\documentclass[a4paper]{IEEEtran}
\usepackage{hyperref} 

\usepackage{xcolor}

\usepackage{graphicx}
\usepackage{tabto}
\usepackage{tabularx}

\graphicspath{ {./images/} }
\definecolor{light-gray}{gray}{0.95}
\newcommand{\code}[1]{\colorbox{light-gray}{\texttt{#1}}}
\usepackage{pdfpages}
\begin{document}
\title{Deep Learning for Speech Emotion Recognition: A CNN Approach Utilizing Mel Spectrograms}
\author{\IEEEauthorblockN{Niketa Penumajji}\\
\IEEEauthorblockA{\textit{Department of Computer Science}\\
\
        {\textit{Kansas State University}}\\
        {\textit{Manhattan, Kansas}}\\
        niketa912@ksu.edu
        }
}
\maketitle
\begin{abstract}
This paper explores the application of Convolutional Neural Networks (CNNs) for classifying emotions in speech through Mel Spectrogram representations of audio files. Traditional methods such as Gaussian Mixture Models and Hidden Markov Models have proven insufficient for practical deployment, prompting a shift towards deep learning techniques. By transforming audio data into a visual format, the CNN model autonomously learns to identify intricate patterns, enhancing classification accuracy. The developed model is integrated into a user-friendly graphical interface, facilitating real-time predictions and potential applications in educational environments. The study aims to advance the understanding of deep learning in speech emotion recognition, assess the model's feasibility, and contribute to the integration of technology in learning contexts.
\end{abstract}

\section{Introduction}

Traditional approaches to the incorporation of technology in learning environments have tended to focus more on interactivity as opposed to analysis, which has resulted in a lack of progress in the areas of performance review, analysis, and overall utility (Cho et al. 2020). This is due in part to the complexity of such environments, with many subjective metrics and mediums that, at first glance, appear to be difficult to study and quantify. In this project, speech emotion has been identified and selected as a suitable metric for analysis. Speech emotion is a language agnostic metric that transcends many of the barriers posed by the more obvious metrics such as spoken / body language or facial expressions. Despite the many nuances of speech, such as tonality or pitch, there are clear patterns that can be derived from studying how emotions are conveyed (Angrick et al. 2019). A spectrogram is a visual representation of the spectrum of frequencies (in this case, the frequencies are converted to the mel scale) across a signal as a function of time. Consider the following mel spectrograms produced from two different male actors saying a short sentence in a blatantly angry and then sad manner.
\\
\begin{figure}[h]
	\centering
	\includegraphics[width=5cm]{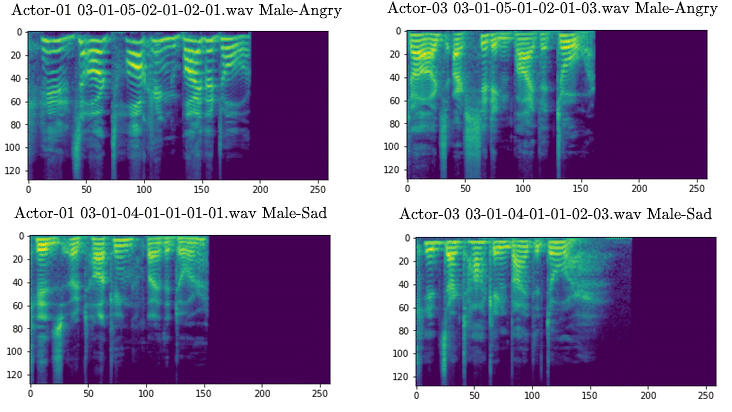}
	\caption{Spectrogram comparison}
\end{figure}
\noindent Even from a macro perspective, it is possible to make out some visible similarities. Take for example the large circular shapes that can be made out from the two angry clips, or the distribution of the yellow (corresponding to pitch across the mel scale) lines in the sad clips. The spectrogram is represented by a matrix of values, which is where the vast, more subtle patterns can be identified and observed - making it an ideal candidate for a CNN approach. Prior to the increased popularity of machine learning, conventional methods for analysing and processing speech involved the usage of Gaussian Mixture Models (GMM) and Hidden Markov Models (HMM), both of which produced results that were simply too inaccurate to ever put into production (Sainath et al., 2013; Venkataramanan \& Rajamohan, 2019). The accessibility of machine learning techniques has lead to novel approaches such as the application of Recurrent Neural Networks (RNN) and CNNs, resulting in improved accuracy and introducing completely different approaches to the problem. In this paper, inspired by the notable success and improvements seen from the application of CNNs in particular, a relatively accurate classification model has been trained by applying CNNs to Mel Spectrogram representations of audio files. The Mel Spectrogram effectively allows me to transform the classification problem from the audio domain into the visual domain, where CNNs have demonstrated to excel. As previously shown in Figure 1, there are many subtle patterns to be found in the data, but it is a difficult task to calculate and label these ourselves manually. Hence, a deep learning approach, where the model learns to recognize and extract features itself, becomes a clear suitable choice. The final trained model has then been deployed into a general purpose graphical tool that can perform predictions across pre-recorded audio files or through recording on the fly. The graphical tool has been purposefully designed with simplicity in mind and with the intention of being easy to use and incorporate into learning environments.
\section{Research Objectives}
\begin{itemize}
\item To research, experiment, and develop an understanding with deep learning using industry standard frameworks.
\item To develop a plausible speech emotion classifier using state of the art techniques.
\item To deploy the speech emotion classification model into an accessible GUI application.
\item To assess the feasibility of such a tool for use in learning environments.
\end{itemize}

\section{Literature Review}

\subsection{Automatic Speech Emotion Recognition Using Machine Learning}
Automatic Speech Emotion Recognition Using Machine Learning explores SER systems, using Mel-frequency cepstrum coefficients and modulation spectral features with feature selection while comparing machine learning approaches. It evaluates an RNN against multivariate linear regression and support vector machines, chosen for their popularity, but omits CNNs, which could have been relevant. Despite this, the paper is well-structured, clear, and accessible to readers with limited prior knowledge. It highlights the advantages of SER in detecting emotional states, especially in learning environments, and details the authors' system, including dataset constraints and algorithm selection. The study concludes that SER performs best with limited data, while RNNs are more effective with larger datasets—insights that are valuable for assessing the effectiveness of my SER system.

\subsection{A CNN-Assisted Enhanced Audio Signal Processing for Speech Emotion Recognition}
Due to the notable omission of a CNN approach in the first reviewed paper, this paper has primarily been selected to gain practical insight into such an approach. This paper explores the CNN approach to SER, focusing on extracting hidden information (Kwon, 2020) and taking a more human-computer interaction (HCI) perspective than the previous paper’s theoretical approach. While it clearly explains its methodology, it assumes prior knowledge of machine learning concepts. The proposed model processes 2D speech spectrograms through convolutional layers and a softmax classifier, requiring less pre-processing than the first paper’s methods. However, it lacks technical depth, with minimal equations and limited training. Despite this, it offers valuable insights, making it a useful comparison to the previous study.
\subsection{Emotion Recognition from Speech}
Emotion Recognition from Speech has been instrumental in shaping the classifier architecture for this project. It compares various industry-standard SER methods, including HMMs, MFCCs, and CNNs, finding that a four-layer model with batch normalization and max pooling achieves the highest accuracy. It also covers data processing techniques like Wiener filtering for noise reduction. Overall, the paper offers valuable architectural insights and comparisons that have greatly influenced this project’s development.
\subsection{Factors Affecting Technology Integration in the Classroom}
This paper outlines five key barriers to technology integration in classrooms: poor infrastructure, inadequate technology, lack of professional development, low self-efficacy, and teacher perceptions. The most relevant to this project are inadequate technology and teacher perceptions. Harrel and Bynum (2018) highlight that teachers assess both the effort required to learn new technology and its practicality before adopting it. Since this project aims to propose a viable application for learning environments, gaining educator buy-in will be essential.
\subsection{Modeling learners' cognitive and affective states to scaffold SRL in open-ended learning environments}
Whilst this paper is not directly related to the project, it offers valuable insights into the relationship between cognitive and affective states, which will aid in the analysis phase. Specifically, the paper highlights a correlation between boredom and delight and academic performance, noting that boredom significantly reduces cognitive and academic performance. With this in mind, focusing on predicting boredom and delight from speech will be key areas for model training.

\section{Methodology}
Selecting the right tools for this project was challenging due to the trade-offs between frameworks. Python was the obvious choice for its extensive machine learning libraries, though a lower-level approach like C++ would have been considered for larger datasets. Initially, PyTorch was chosen for its intuitive syntax, but its unclear documentation made implementation difficult. Switching to Keras and TensorFlow proved more successful due to TensorFlow’s superior documentation, enabling efficient model development with minimal debugging. For audio processing, Librosa was used for its well-documented and concise handling of complex transformations.

\section{Implementation}
The development of this project can be broken down into 5 key stages: dataset selection and comprehension, data processing, model architecture and training, testing, and finally, GUI development.
\subsection{Foreword}
A lot of the time spent on this project was spent prior to the key stages on personal development; this involved developing a theoretical and practical understanding of machine learning techniques. In particular, two LinkedIn Learning courses by Fernandes (2019) and Geitgey (2017) both played a large role in understanding how the technology worked at both the practical and theoretical levels. Once a core understanding was developed, consistent literature review of technical papers on speech emotion recognition helped to further shape the understanding towards the specific subject matter. 
\subsection{Dataset Selection and Comprehension}
Prior to embarking on any practical work for this project, 
careful consideration was given to the dataset for training and testing the speech emotion classifier. Collecting original data was initially considered but ruled out due to the global pandemic. Available options were limited, with the Ryerson Audio-Visual Database of Emotional Speech and Song (RAVDESS) being the best choice. This open-source dataset contains 1,440 labeled audio clips, each expressing a distinct emotion: calm, happy, sad, angry, fearful, surprised, or disgusted.
\\
\indent Using RAVDESS meant the project’s emotion scope was predefined, making prior research on selecting optimal emotions for learning environments redundant. However, the dataset covered general emotions initially considered, and later findings showed that more complex emotions could be inferred from broad parent classes and multi-class weighted predictions as shown below in Figure 2.
\\
\begin{figure}[h]
	\centering
	\includegraphics[width=5cm]{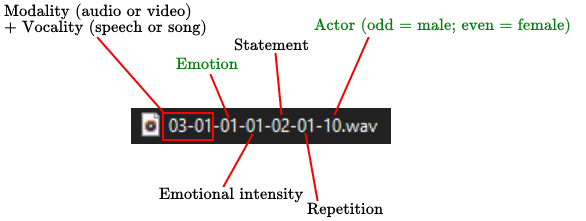}
	\caption{An example audio file from the dataset; only the sections marked in green (emotion and gender) need to be taken into consideration.}
\end{figure}
Since the dataset is well-structured and labeled, the only task was extracting key labels, achieved using the \code{get\_label\_RAVDESS} helper function (see Appendix). While emotional intensity was initially considered, the dataset’s size made additional categories impractical. Initially, all labeled emotions were included, but this led to poorer results due to an uneven emotion distribution. Further analysis showed that minority emotions like calm and surprise could be merged to strengthen neutral and happy classifications, improving overall model performance. 
\\
\begin{table*}[h]
\centering
\caption{Part 1 of the label conversion table}
\begin{tabularx}{\textwidth}{ |X|X| }
\hline 
  \textbf{Original Label}  & \textbf{Converted Label} \\
  \hline
  Neutral & Neutral \\
  \hline
  Calm & \textbf{Neutral} \\
  \hline
  Happy & Happy \\
  \hline
  Sad & Sad \\
  \hline
\end{tabularx}
\end{table*}

\subsection{Data Processing}
Since the dataset comes in the form of labeled audio (.wav) files, the main task is to convert these to spectrograms, in particular mel spectrograms, that can later be used to train the model. Using the Mel scale as opposed to the Hertz scale is important when working with human audio; the Mel scale is a scale derived from tests with human listeners, such that the scale of pitches is judged to be equal in distance from one another.
\\
\indent The first task is converting the audio frequency from the hertz scale to the mel scale. This is given by the equation $m = 2595 \log_{10}\left(1 + \frac{Hz}{700}\right)$. Once this conversion is made, a mel spectrogram can be obtained from an audio signal by computing the squared magnitude of the Short-Time Fourier Transform (STFT) applied to a signal, substituting the frequency of the signal for the mel scale conversion. Finally, the mel spectrogram is scaled to decibel units according to $10 \times \log_{10}(mel \, spectrogram)$. This, in turn, corresponds to a weighted matrix, as in Figure 4 below.
\begin{figure}[h]
	\centering
	\includegraphics[width=5cm]{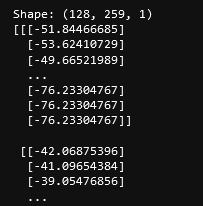}
	\caption{A log-scaled mel spectrogram obtained from an audio file from the dataset}
\end{figure}

\begin{figure}[h]
	\centering
	\includegraphics[width=5cm]{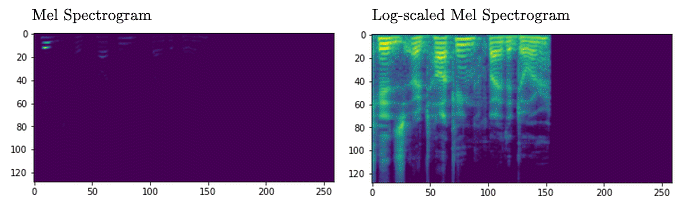}
	\caption{Figure illustrates the importance of decibel conversion}
\end{figure}
\indent The last step of the data processing stage was creating the testing and training data by applying the labeling and spectrogram transformation functions to each audio file. The testing data was selected through a random sampling ($n = 180$) of assorted audio files containing all emotions in scope and all actors except actor 24. Actor 24 was excluded in an attempt to introduce some completely unknown data into the testing dataset.

\subsection{Model Architecture \& Training}
The model was built through research and experimentation, with particular focus on the convolutional neural network (CNN) approach, which showed the best results in the literature, notably in Venkataramanan and Rajomohan (2019). CNNs analyze discrete features within tensors, which the spectrogram (a matrix representation of audio) easily converts into. This made CNNs suitable for the project, transitioning the problem from audio to visual data.
\\
\indent After finalizing the design, the model was trained using various batch and epoch sizes. The optimal results came from 125 epochs with a batch size of 16, balancing optimization and avoiding overfitting. The architecture remained consistent throughout the tests, comprising 4 2D convolutional layers with max pooling, batch normalization, dropout regularization, and ELU activation. The Adam optimizer was used for its adaptive learning rate, and categorical cross-entropy was chosen as the loss function due to the multi-class classification goal.
\begin{figure}[h]
	\centering
	\includegraphics[width=5cm]{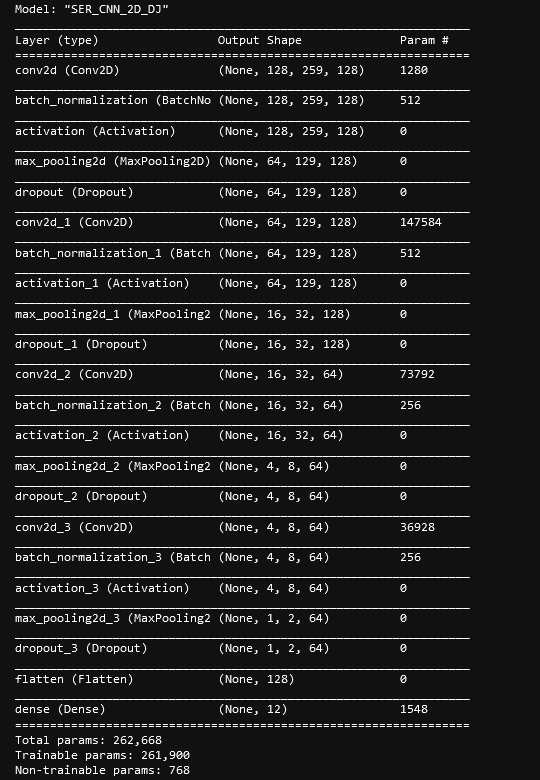}
	\caption{Summary of the complete speech emotion recognition model}
\end{figure}

\indent The training steps and code can be viewed from the SER Classifier Jupyter Notebook with link in the appendix.
\subsection{Testing}
After training the final model, it was tested on a blind dataset of 180 predictions, achieving a categorical accuracy of 68.88\%. While this accuracy may be inflated due to the dataset’s similarities (all North American actors speaking English), it is still impressive given the limited training data and the use of actors not included in the training set.

\begin{table*}[ht]
\centering
\caption{Part 1 of the table: Actual values and predictions}
\begin{tabularx}{\textwidth}{ |X|X| }
\hline 
  \textbf{Actual value} & \textbf{Prediction} \\
  \hline
  22. female sad & 22. female neutral \\
  \hline
  28. female angry & 28. female neutral \\
  \hline
  45. female disgust & 45. male sad \\
  \hline
  72. male happy & 72. female happy \\
  \hline
\end{tabularx}
\end{table*}
Figure 7 - a sample of the incorrect predictions, showing reasonable mistakes (i.e., emotional similarity or right emotion; wrong gender). \textit{Values taken from SER\_Classifier notebook}.
\\
\indent Next, the model was tested on unique audio from myself, family, and friends. Surprisingly, it performed well, especially with negative emotions. For example, it correctly predicted male anger with over 90\% accuracy, often distinguishing it from other emotions like male disgust, female anger, and male sadness.

\begin{figure}[h]
\centering
\includegraphics[width=7cm, height=6cm]{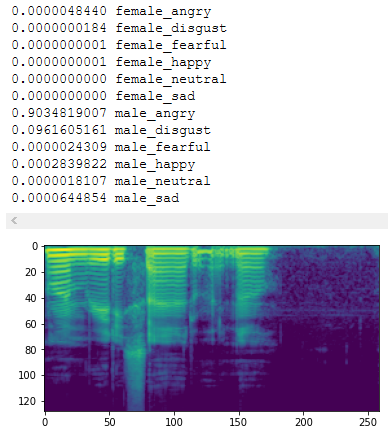}
\caption{Output predictions using an audio clip of myself talking angrily, available through the appendix.}
\label{fig:personal_prediction}
\end{figure}

\indent An interesting test involved a friend with Asperger's syndrome, who struggles with recognizing emotions. While the model's accuracy seemed initially low, further analysis revealed that her own perception of emotions was misaligned with the model's predictions, which were actually more accurate.

\begin{figure}[h]
\centering
\includegraphics[width=0.4\textwidth]{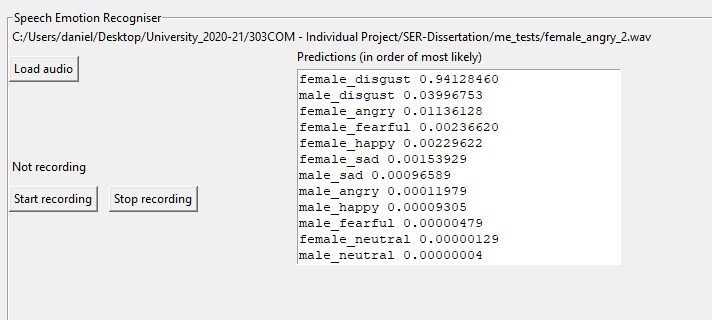}
\caption{An example of one of the tests on a friend with autism who had herself misclassified the audio (and a preview of the GUI tool).}
\label{fig:test}
\end{figure}

\indent Finally, the model was tested on German and Swiss German audio, where it performed well in predicting anger, sadness, and disgust. However, it made some errors with positive emotions. In all cases of failure, the target emotion remained within the top 5 predicted classes, demonstrating the model's robustness.
\subsection{GUI Development}
Once the model was developed and tested, the final step was to deploy it into a user-friendly tool. The goal was for users with no technical background to easily obtain results without writing code or modifying files. The tool’s core functionality was simple: it needed to analyze pre-recorded audio and allow users to record and analyze audio directly, with clear results displayed. The tool was designed for accessibility, focusing on simplicity and efficiency. Built natively in Python using Tkinter with minimal libraries and graphics, the tool is shown in Figure 10, with a working example in Figure 9.

\begin{figure}[h]
	\centering
	\includegraphics[width=5cm]{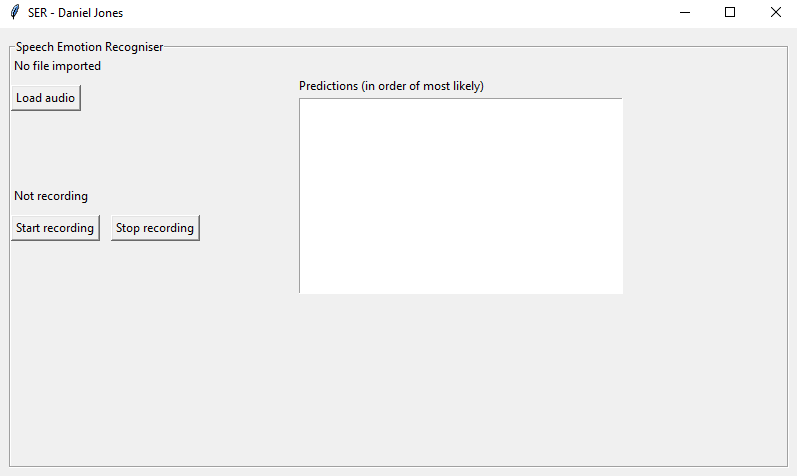}
	\caption{The tool, upon execution, shows the two key functions as buttons}
\end{figure}

\section{Evaluation}
The final result of this project is a promising speech emotion recognition classifier, demonstrated by its tested accuracy on both the test dataset and through personal tests, as well as a fully functional tool for end users. In this section, I will evaluate the successes and failures of the project in relation to the original research objectives outlined in the project proposal under the ``Primary Research Plan''.
\subsection{Research Objectives}
The primary research objective was to build a speech emotion classifier based on current literature. Through various tests and experiments, I successfully developed and trained a viable model, utilizing methods that deliver competitive results. This was achieved through ongoing literature review and personal research into machine learning best practices. The final model and associated work serve as a clear approach to tackling the problem, highlighting both successes and areas for improvement. 
\\
\indent The second objective was to improve the model through formal tests and research in learning environments. Unfortunately, this objective could not be fully realized due to logistical limitations. However, the model is in a solid state, and with additional data collection, processing, and retraining, it can easily be further developed.
\\
\indent The final objective, to create a tool for use in learning environments, has been achieved. A lightweight, user-friendly GUI has been developed, yielding quick results with minimal dependencies. Testing with an autism spectrum user showed that the tool provides useful insights. Given the simplicity and lightweight design of the GUI, further improvements would require model adjustments or real-world feedback from learning environments.
\\
\indent Overall, significant progress has been made in achieving the research objectives. While not all objectives were perfectly met, the project is at a promising stage, and I intend to continue its development in future projects. 

\section{Conclusion}
The project was structured around building a speech emotion recognition classifier, creating an accessible tool, and testing it in learning environments. While the third stage was not completed due to the pandemic and time constraints, the project has been a valuable learning experience. The final product offers useful insights, particularly from testing with a friend with autism. The model architecture and trained model provide a solid foundation for future work. Although the learning environment testing wasn't conducted, the tool is well-documented and can easily be continued. Overall, the project has been a significant personal success.
\
\bibliographystyle{IEEEtran}

\end{document}